\documentclass[modern]{aastex631}

\begin{document}

\title{Anomalous flux event in the TESS Sector 43 light curve of the white dwarf photometric standard HZ 4 was caused by a passing asteroid}

\author[0000-0002-0656-032X]{Keaton J.\ Bell}
\affiliation{Department of Physics, Queens College, City University of New York, Flushing, NY-11367, USA}
\email{Keaton.Bell@qc.cuny.edu}

\author[0000-0002-2564-8116]{David Ardila}
\affiliation{Jet Propulsion Laboratory, California Institute of Technology, USA}

\author[0009-0004-6219-096X]{Alexandra Frymire}
\affiliation{Vanderbilt University, USA}

\begin{abstract}
\citet{2023RNAAS...7...94F} reported an anomalous flux variation in the Transiting Exoplanet Survey Satellite (TESS) Sector 43 light curve of the white dwarf HZ~4.
We show that this flux variation was caused by the main-belt asteroid 4382 Stravinsky traversing the nearby TESS pixels, and it is therefore not a cause for concern regarding the continued use of HZ~4 as a photometric standard star. 

\end{abstract}


\section*{} 

Precise space-based, time-domain photometric data from \textit{Kepler} \citep{Kepler}, K2 \citep{K2}, and the Transiting Exoplanet Survey Satellite \citep[TESS;][]{TESS} have been used to validate the constant brightness of white dwarf stars that are used as photometric standards \citep{2017MNRAS.468.1946H,2022AJ....163..136M}. The hydrogen-atmosphere white dwarf HZ~4 is a commonly referenced photometric standard star included in the CALSPEC\footnote{\url{https://www.stsci.edu/hst/instrumentation/reference-data-for-calibration-and-tools/astronomical-catalogs/calspec}} archive of flux standards used by the Hubble Space Telescope \citep{2014PASP..126..711B}. \citet{2023RNAAS...7...94F} recently reported a dramatic flux variation observed in the TESS Sector 43 light curve of HZ~4. The event lasted 28.8 hours, involving a flux increase of 43\%, bracketed by flux decreases of 11\%, which called into question the reliability of HZ~4 as a photometric standard.

\begin{figure}[t]
\centering{
\includegraphics[width=1\textwidth,angle=0]{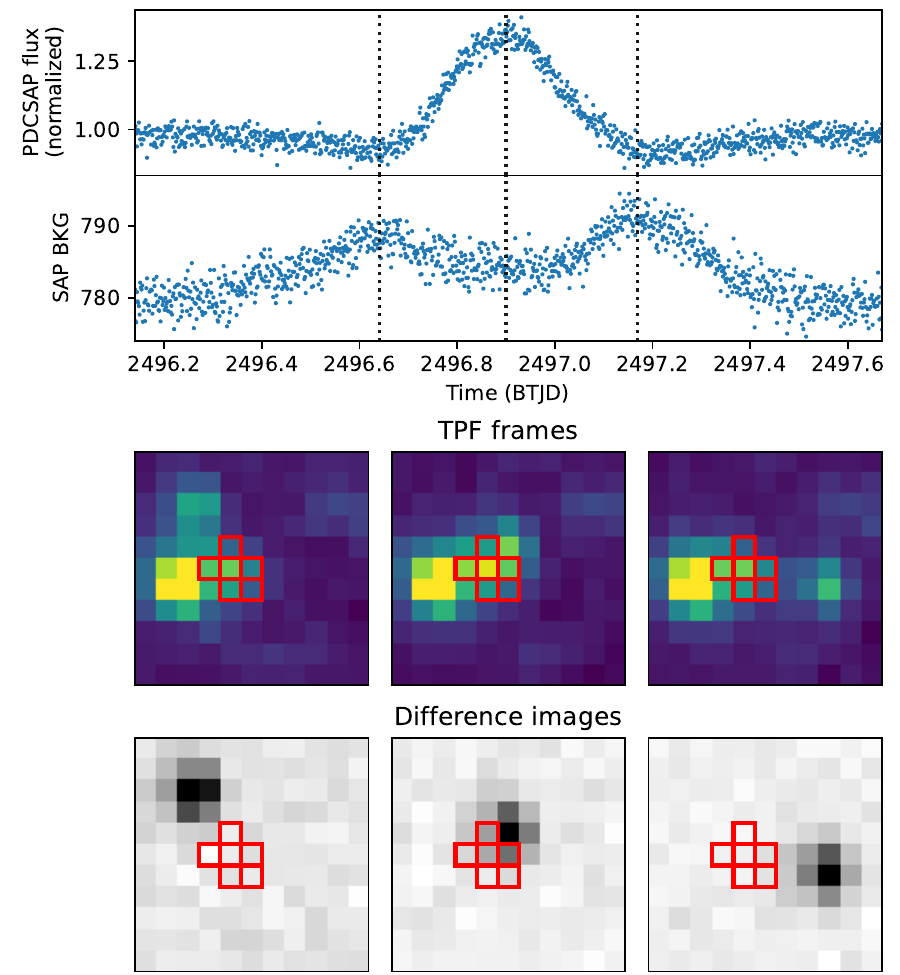}
}
\vspace*{-0.5em}
	\caption{Evidence that the anomalous variations in the TESS Sector 43 light curve of HZ~4 are caused by a passing asteroid. 
 The top panel shows the median-normalized, 120-second cadence, pre-search data conditioned simple aperture photometry (PDCSAP) light curve centered on the anomalous flux event reported by \citet{2023RNAAS...7...94F}. Dotted vertical lines mark times near the maximum increase and the two surrounding decreases of flux in the pipeline-produced light curve during the anomaly. 
 The second panel shows the estimated background flux that was subtracted from the flux in the aperture ({\tt sap\_bkg}), which shows increases that correspond to the decreases of the PDCSAP light curve. 
 The third row shows frames from the target pixel file (TPF) images corresponding to the dotted lines in the upper panels, with the pixels included in the photometric aperture used to extract the light curve outlined in red. The bottom panels show difference images produced by subtracting the median TPF image from these same TPF frames. A moving object is clearly passing across the pixels during the flux anomaly, grazing the photometric aperture.
	}
\label{fig1}
\end{figure}

TESS has a large plate scale of 21\arcsec\,pixel$^{-1}$, resulting in photometric apertures often capturing a combination of flux from multiple sources. When analyzing TESS light curves, there is considerable risk of misattributing brightness variations from a contaminating source to the intended target of study \citep[e.g.,][]{2023AJ....165..239P}. These contaminants can include moving Solar System objects such as asteroids, which TESS is sensitive to detecting \citep{2018PASP..130k4503P}.

In the case of HZ~4, the variations reported by \citet{2023RNAAS...7...94F} in TESS Sector 43 were caused by an asteroid in the Solar System passing near the photometric aperture. This is demonstrated in Figure~\ref{fig1}, where an asteroid is visible moving across the TESS target pixel file (TPF) images during the flux anomaly (in both raw and difference images). The peak brightness enhancement occurs as the asteroid grazes the photometric aperture, and the surrounding flux decreases occur when the asteroid is in the TPF outside of the aperture, affecting the local background subtraction. The times when the light curve appears dimmer correspond to increases in the subtracted background level that are recorded in the \verb+sap_bkg+ column of the light curve file. 

Due to the high risk of source contamination in TESS data, it is important to analyze the TESS images in addition to the light curve data products. There are a few Python packages that are helpful for identifying contamination in the pixel data from passing Solar System objects. The {\tt lightkurve} package \citep{lightkurve} provides a suite of functions that are invaluable for analyzing TESS data. The {\tt lightkurve.TessTargetPixelFile.interact} function, in particular, can be used to interactively inspect the TESS TPF frames for moving objects. The {\tt lightkurve.LightCurve.query\_solar\_system\_objects} function can be used to identify the specific Solar System object (if known) that is moving through the TPF; this reveals the offending object in the case of HZ~4 to be the main-belt asteroid 4382 Stravinsky. The {\tt k2flux} package \citep{k2flux} is also useful for creating animations of the TPF frames, where asteroids traversing the image may be visible.

Since the flux variations identified in the TESS light curve of HZ~4 by \citet{2023RNAAS...7...94F} are not intrinsic to the star, they do not disqualify HZ~4 as a reliable photometric standard.

\begin{acknowledgments}
This paper includes data collected with the TESS mission, obtained from the MAST data archive at the Space Telescope Science Institute (STScI); the data is available at \dataset[10.17909/02qr-hh98]{\doi{10.17909/02qr-hh98}}. Funding for the TESS mission is provided by the NASA Explorer Program. STScI is operated by the Association of Universities for Research in Astronomy, Inc., under NASA contract NAS 5–26555. This research made use of Lightkurve, a Python package for Kepler and TESS data analysis \citep{lightkurve}. This work made use of Astropy:\footnote{http://www.astropy.org} a community-developed core Python package and an ecosystem of tools and resources for astronomy \citep{astropy:2013, astropy:2018, astropy:2022}. This research has made use of IMCCE's SkyBoT VO tool. The research was carried out in part at the Jet Propulsion Laboratory (JPL), California Institute of Technology, under a contract with the National Aeronautics and Space Administration (80NM0018D0004). A.F.\ thanks the Center for Academic Partnerships at JPL for their support. Thank you to Courtney Crawford and Tim White for direction in how to identify the offending asteroid in this case.

\end{acknowledgments}

\bibliography{bibliography}{}
\bibliographystyle{aasjournal}

\end{document}